# Tunable parametric amplification of a graphene nanomechanical resonator in the nonlinear regime

Zi-Jia Su[1,2], Yue Ying[1,2], Xiang-Xiang Song[1,2*], Zhuo-Zhi Zhang[1,2], Qing-Hang Zhang[1,2], Gang Cao[1,2], Hai-Ou Li[1,2], Guang-Can Guo[1,2], and Guo-Ping Guo[1,2,3†]

1. CAS Key Laboratory of Quantum Information, University of Science and Technology of China, Hefei, Anhui 230026, China
2. CAS Center for Excellence in Quantum Information and Quantum Physics, University of Science and Technology of China, Hefei, Anhui 230026, China
3. Origin Quantum Computing Company Limited, Hefei, Anhui 230088, China

Author to whom correspondence should be addressed:
*X.-X. Song (songxx90@ustc.edu.cn) or †G.-P. G. (gpguo@ustc.edu.cn).

# Abstract

Parametric amplification is widely used in nanoelectro-mechanical systems to enhance the transduced mechanical signals. Although parametric amplification has been studied in different mechanical resonator systems, the nonlinear dynamics involved receives less attention. Taking advantage of the excellent electrical and mechanical properties of graphene, we demonstrate electrical tunable parametric amplification using a doubly clamped graphene nanomechanical resonator. By applying external microwave pumping with twice the resonant frequency, we investigate parametric amplification in the nonlinear regime. We experimentally show that the extracted coefficient of the nonlinear Duffing force $\alpha$ and the nonlinear damping coefficient $\eta$ vary as a function of external pumping power, indicating the influence of higher-order nonlinearity beyond the Duffing ($\sim x^3$) and van der Pol ($\sim x^2 \frac{dx}{dt}$) types in our device. Even when the higher-order nonlinearity is involved, parametric amplification still can be achieved in the nonlinear regime. The parametric gain increases and shows a tendency of saturation with increasing external pumping power. Further, the parametric gain can be electrically tuned by the gate voltage with a maximum gain of 10.2 dB achieved at the gate voltage of 19 V. Our results will benefit studies on nonlinear dynamics, especially nonlinear damping in graphene nanomechanical resonators that has been debated in the community over past decade.

# 1. Introduction

Graphene is considered a promising material for nanoelectro-mechanical systems applications because of its excellent electrical[1,2] and mechanical properties[3,4]. Taking advantage of high resonant frequency[5], high quality factor[6] and good electrical tunability[7], graphene-based nanomechanical resonators are widely used in ultrasensitive detection[8-13], cavity optomechanics[14-18] and especially nonlinear physics research[19-21], including Duffing nonlinearity[22] and nonlinear damping[23,24]. In these applications, to enhance the transduced mechanical signals, an effective solution is parametric amplification, which is achieved by using the energy from an external pump. One of the most conventional methods is pumping at a frequency twice the resonant frequency of the nanomechanical resonator, which modulates the spring constant of the resonator[25]. Such a technique is widely used in studies on nanomechanical resonators[25-31]. Considering the debating on the mechanism of nonlinearity, especially the nonlinear damping of graphene nanomechanical resonators over past decade[6,23,24,32-34], although parametric amplification has been observed in graphene nanomechanical resonators[28,33], more efforts are needed to understand the nonlinear dynamics involved in parametric amplification.

Here, we use a doubly clamped graphene nanomechanical resonator[35] to demonstrate electrical tunable parametric amplification of mechanical motion. We can drive the graphene nanomechanical resonator in the nonlinear regime and investigate nonlinear dynamics involved in parametric amplification. The absolute values of extracted coefficient of the nonlinear Duffing force $\alpha$ and the nonlinear damping coefficient $\eta$ increase with increased power of external pumping, suggesting higher-order nonlinearity[36] beyond the Duffing ($\sim x^3$) and van der Pol ($\sim x^2 \frac{dx}{dt}$) types exist in our device. In the nonlinear regime, we demonstrate that the parametric gain increases and shows a tendency of saturation when increasing the power of external pumping. Further, the parametric gain can be electrically tuned by varying gate voltage. A maximum gain as high as 10.2 dB can be achieved at the gate voltage of 19 V in the nonlinear regime. Our results not only provide the possibility of applications for ultrasensitive detection and sensing, but also benefit the study of nonlinear dynamics in graphene nanomechanical resonators.

# 2. Device fabrication

Figure 1(a) shows a scanning electron microscopy (SEM) image of the device investigated in the experiment. A few-layer graphene flake is suspended over a prepatterned trench, serving as a nanomechanical resonator. A schematic of the cross-section is shown in Fig. 1(b). First, a layer of $SiN_x$ (50 nm) is deposited via low-pressure chemical vapor deposition (LPCVD) on a silicon oxide layer (with a thickness of 300 nm), which covers a highly resistive silicon wafer. After electron beam lithography (EBL), a trench is etched through the $SiN_x$ layer by reactive ion etching, followed by dipping in hydrofluoric acid. The total etching depth is approximately 170 nm. The width of the trench is designed to be 2 μm. After a second EBL step, 3 nm titanium and 20 nm gold are evaporated onto the wafer. Two contacts with widths of ~1 μm are defined as the source and drain, labeled as S and D, respectively, in Fig. 1(a). A bottom electrode is defined as a gate (labeled as G in Fig. 1(a)) for electrical tuning. Finally, a graphene ribbon, exfoliated on a polydimethylsiloxane (PDMS) stamp, is aligned and transferred above the trench[37]. The suspended graphene has a length of 2 μm and a width of 400 nm. All measurements were carried out at ~250 mK and a pressure below $10^{-7}$ Torr.

## 3. Results and discussion

We characterize the nanomechanical resonator using the one-source frequency modulation (FM) technique[38]. The graphene resonator is actuated by the FM signal with an amplitude $\tilde{V}_f^{FM}$ at the driving frequency $f$. The FM signal has the form[38]:

$$\tilde{V}_f^{FM}(t) = V_d \cos(2\pi f t + \left(\frac{f_\Delta}{f_L}\right) \sin(2\pi f_L t)) \tag{1}$$

where $V_d$ is the amplitude of the driving voltage, $f_\Delta$ is the deviation frequency (typically 43 kHz in the experiment), and $f_L$ is the modulation frequency (typically 1.03 kHz). The mixing current $I_{mix}$ at frequency $f_L$ is detected by a lock-in amplifier at the drain electrode. A dc gate voltage ($V_g$) is applied to the gate to adjust the strain of the resonator, as illustrated in Fig. 1(a).

Figure 1(c) shows the spectrum of the device when varying the gate voltage $V_g$. As $V_g$ increases, the resonant frequency increases, indicating that the strain induced by electrostatic force becomes larger when tuning $V_g$ to a more positive value. The parabolic shape suggests a low built-in tension is obtained during fabrication[8]. By fitting the resonant frequency as a function of $V_g$, we can extract the effective mass[23] $m = 5.7 \times 10^{-18}$ kg, as shown in Fig. 1(d).

To realize parametric amplification, continuous external pumping at a frequency of 2$f$ (labeled as $\tilde{V}_{2f}$) is applied to the gate through a bias-tee, as shown in Fig. 1(a). We define the power of the FM signal $\tilde{V}_f^{FM}$ as $P_f^{FM}$ and the power of external pumping $\tilde{V}_{2f}$ as $P_{2f}$. Here, $P_f^{FM}$ is minimized to −62 dBm to decrease the influence of $\tilde{V}_f^{FM}$. Figure 2(a) shows the mixing current as a function of $f$ and $P_{2f}$ ($V_g$ is set to 16.5 V). The mixing current at resonance increases when increasing $P_{2f}$. We plot the mixing current as a function of driving frequency $f$ at two typical pumping powers $P_{2f}$=−56 dBm and $P_{2f}$=−46 dBm in Fig. 2(b) and Fig. 2(c), respectively. As shown in Fig. 2(b), when $P_{2f}$=−56 dBm, the response (blue open circles) shows a typical line shape of mixing current without considering nonlinearity obtained using the FM technique[38], suggesting that the nanomechanical resonator is in the linear regime. In the linear regime, we have:

$$x_0 = \frac{F_{drive}^{FM}}{m(\omega_0^2 - \omega^2) + i\omega\gamma} \quad (2)$$

where $\omega = 2\pi f$, $\omega_0 = 2\pi f_0$, with $f$ is the driving frequency, $f_0$ is the resonant frequency, $\gamma$ is the linear damping coefficient. Here the influence of driving force is considered mainly from $\tilde{V}_f^{FM}$ and thus can be estimated using $F_{drive}^{FM} = 1.25 C' V_g V_p^{AC} \approx -0.9$ pN, with $C' \approx -2.45 \times 10^{-10}$ F/m, is the derivative of the capacitance estimated using a parallel-plate capacitor model, $V_p^{AC}$= 0.178 mV is the amplitude of the FM signal voltage. Note that in our experiment, there is a background of mixing current. We use $I_{mix} \propto \partial Re[x_0]/\partial f$ to fit the experimental data, where $Re[x_0]$ is the real part of the vibrational amplitude:

$$I_{mix} \propto F_{drive}^{FM} m \frac{\omega[m^2(\omega_0^2 - \omega^2)^2 - \gamma^2 \omega_0^2]}{[m^2(\omega_0^2 - \omega^2)^2 + (\gamma\omega)^2]^2} \quad (3)$$

As shown in Fig. 2(b), the experimental data in the linear regime (blue open circles) can be fitted well using Eq. (3). The linear damping coefficient $\gamma = 7.66 \times 10^{-12}$ kg·s$^{-1}$ can be obtained from fitting (shown by the black solid curve). We assume $\gamma$ remains unchanged in both linear regime and nonlinear regime described later.

When further increasing $P_{2f}$, the mixing current develops an asymmetry (red open circles in Fig. 2(c)) by parametric pumping, suggesting that the nanomechanical resonator is driven into the nonlinear regime. The results can be quantitatively understood using the model written as[23]:

$$m\frac{d^2 x}{dt^2} = -kx - \gamma\frac{dx}{dt} - \eta x^2 \frac{dx}{dt} - \alpha x^3 + F_{drive}^{eq} \quad (4)$$

where $m$ is the effective mass of the nanomechanical resonator, $k$ is the spring constant, and $\alpha$

is the coefficient of the nonlinear Duffing force. In our model, we consider two ways of dissipation: linear term $\gamma dx/dt$ and nonlinear term[39] $\eta x^2 dx/dt$. Here, different from $F_{drive}^{FM}$, $F_{drive}^{eq}$ is considered an equivalent driving force consists not only the direct driving $\tilde{V}_f^{FM}$ but also the influence of parametric pumping $\tilde{V}_{2f}$. In the nonlinear regime, the vibrational amplitude $x_0$ and phase $\phi$ can be calculated using[23]:

$$x_0^2 = \frac{(F_{drive}^{eq}/4\pi f_0)^2}{(m(2\pi f - 2\pi f_0) - \frac{3}{16}\frac{\alpha}{\pi f_0}x_0^2)^2 + (\frac{1}{2}\gamma + \frac{1}{8}\eta x_0^2)^2} \quad (5)$$

$$\tan(\phi) = \frac{\gamma/2 + \eta x_0^2/8}{m(2\pi f - 2\pi f_0) - \frac{3}{16}\frac{\alpha}{\pi f_0}x_0^2} \quad (6)$$

Solving Eqs. (5)-(6), we can calculate $Re[x_0] = x_0 \cos(\phi + \theta)$ and use $I_{mix} \propto \partial Re[x_0]/\partial f$ to fit our experimental data in the nonlinear regime. Here an additional phase $\theta$ induced by the electrical circuit is considered[40] and is set to be a fitting parameter. As shown in Fig. 2(c), the experimental data in the nonlinear regime (red open circles) can be fitted well using our model.

From fitting, the values of $\alpha$, $\eta$ and $|F_{drive}^{eq}|$ can be extracted. We plot the extracted $\alpha$, $\eta$ and $|F_{drive}^{eq}|$ as a function of $P_{2f}$ in the nonlinear regime, as shown in Fig. 3(a)-(c). When the pumping power $P_{2f}$ increases, the absolute value of $\alpha$ and $\eta$ increases with $P_{2f}$, as shown in Fig. 3(a)-(b). Meanwhile, $|F_{drive}^{eq}|$ also increases (see Fig. 3(c)). The estimated $|\alpha| \sim 10^{15}\ kg \cdot m^{-2} s^{-2}$ is comparable to the result obtained from a graphene nanomechanical resonator[19,22], while the estimated $\eta \sim 10^6\ kg \cdot m^{-2} s^{-1}$ is one order of magnitude larger[19].

Moreover, we can quantitatively characterize the effect of parametric amplification by extracting the gain of parametric amplification. The amplification gain is defined as $Gain = 20 \cdot \lg(x_0^{pumped}/x_0^{unpumped})$ with $x_0^{unpumped}$ is the vibrational amplitude extracted when parametric pumping is not significant ($P_{2f}= -61$ dBm) and $x_0^{pumped}$ is the vibrational amplitude when the influence of parametric pumping is pronounced. We plot $Gain$ (red solid squares) at $V_g$=16.5 V as a function of $P_{2f}$ in Fig. 3(d). As shown in Fig. 3(d), $Gain$ increases with increasing $P_{2f}$ and shows a tendency of saturation. This is probably due to competition between parametric pumping and more pronounced influence of nonlinearity at higher external pumping power, as the nonlinear coefficient $\eta$ is found to increase sharply with $P_{2f}$ (see Fig3. (b)). The maximum $Gain$ is estimated to be ~8.5 dB at $V_g$=16.5 V. It is worth noting that parametric amplification is sensitive to the phase $\varphi$ between driving and parametric pumping[25]. Here in our measurement setup, the FM technique will introduce a time varying phase lag, and hence, the overall gain is an average result due to varying phases[30]:

$$Gain = \frac{1}{2\pi}\int_0^{2\pi} G(\varphi)d\varphi = \frac{1}{2\pi}\int_0^{2\pi}[\frac{cos^2\varphi}{(1+V_p/V_t)^2} + \frac{sin^2\varphi}{(1-V_p/V_t)^2}]^{1/2}d\varphi = (1-(V_p/V_t)^2)^{-1} \quad (7)$$

where $V_t$ is the threshold voltage determined by the system parameters, and $V_p$ is the pumping voltage for parametric amplification. We fitted the data using Eq. (7), as shown by the blue curve in Fig. 3(d). Our model fits well when $P_{2f}$ is low while deviates at larger $P_{2f}$ where the influence of nonlinearity becomes more pronounced.

In doubly clamped graphene nanomechanical resonator, the nonlinear coefficients $\alpha$ and $\eta$ should be independent of vibrational amplitude[36] (thus $P_{2f}$). However, as shown in Fig. 3(a), the extracted $|\alpha|$ increases with increasing $P_{2f}$, suggesting that $|\alpha|$ is a function of vibrational amplitude as $\alpha(x)$. We further plot $|\alpha|$ as a function of vibrational amplitude $x_0$, as shown in Fig. 4(a). Indeed, $|\alpha|$ increases with $x_0$ almost linearly. Thus the term of $\alpha(x)x^3$ is beyond the Duffing nonlinearity ($\sim x^3$), which indicates that higher-order nonlinearity[36] (such as $\sim x^4$) is involved in our experiment. Similarly, as shown in Fig. 4(b), $\eta$ also increases with $x_0$, in a more complicated way however. The blue solid line in Fig. 4(b) is a linear fitting for the data when $x_0$ is smaller than 0.53 nm. When $x_0$ is driven to a larger value, $\eta$ shows a faster increasing than linear dependency. Nevertheless, since $\eta$ is a function of vibrational amplitude as $\eta(x)$, $\eta(x)x^2\frac{dx}{dt}$ is also beyond the commonly observed van der Pol nonlinearity ($\sim x^2\frac{dx}{dt}$), while detailed form of such higher-order nonlinear damping needs further investigation. It is worth noting that although the influence of higher-order nonlinearity is pronounced in our device, the parametric amplification can still be achieved.

Finally, we investigate parametric amplification in the nonlinear regime at different gate voltage $V_g$. To compare the gain of parametric amplification at different gate voltage, we keep $P_f^{FM}$ to be constant as $-56$ dBm and $P_{2f} = -44$ dBm. The $Gain = x_0^{pumped}/x_0^{unpumped}$ as a function of gate voltage $V_g$ is shown in Fig. 5. The $Gain$ can be electrically tuned by $V_g$. The maximum $Gain$ of 10.2 dB is obtained at $V_g$=19 V, which is comparable with results observed in graphene drum resonators[28].

## 4. Conclusions

In summary, we demonstrate electrically tunable parametric amplification in a doubly clamped graphene nanomechanical resonator using the FM mixing technique. The extracted Duffing nonlinear coefficient $\alpha$ and nonlinear damping coefficient $\eta$ is found to be functions of external pumping power $P_{2f}$, suggesting higher-order nonlinearity exists in our graphene nanomechanical resonator. The nonlinearity in nanomechanical resonators may be related to phonon tunneling, sliding at contacts, nonlinearity in phonon-phonon interactions, or contamination in combination with geometrical nonlinearity[23]. The detailed origin and model of this higher-order nonlinearity need to be further investigated. Even when the influence of higher-order nonlinearity is pronounced, the parametric amplification can still be achieved. Moreover, the parametric gain can be electrically tuned by gate voltage. The largest gain of 10.2 dB is observed in the nonlinear regime. Our design of a graphene nanomechanical resonator can be further extended[41,42], which will provide a scalable architecture for potential applications of tunable parametric amplification. The ability to amplify the vibrational amplitude in the nonlinear regime makes graphene nanomechanical resonators attractive for a wide variety of sensing applications. Additionally, our results will benefit studies of nonlinear dynamics in graphene nanomechanical resonators.

**Acknowledgements**

This work was supported by the National Key Research and Development Program of China (Grant No. 2016YFA0301700), the National Natural Science Foundation of China (Grants Nos. 11625419, 61674132, 11674300, 61904171, and 11904351), the China Postdoctoral Science Foundation (Grant No. BX20180295), and the Anhui Initiative in Quantum Information Technologies (Grants No. AHY080000). This work was partially carried out at the USTC Center for Micro and Nanoscale Research and Fabrication.

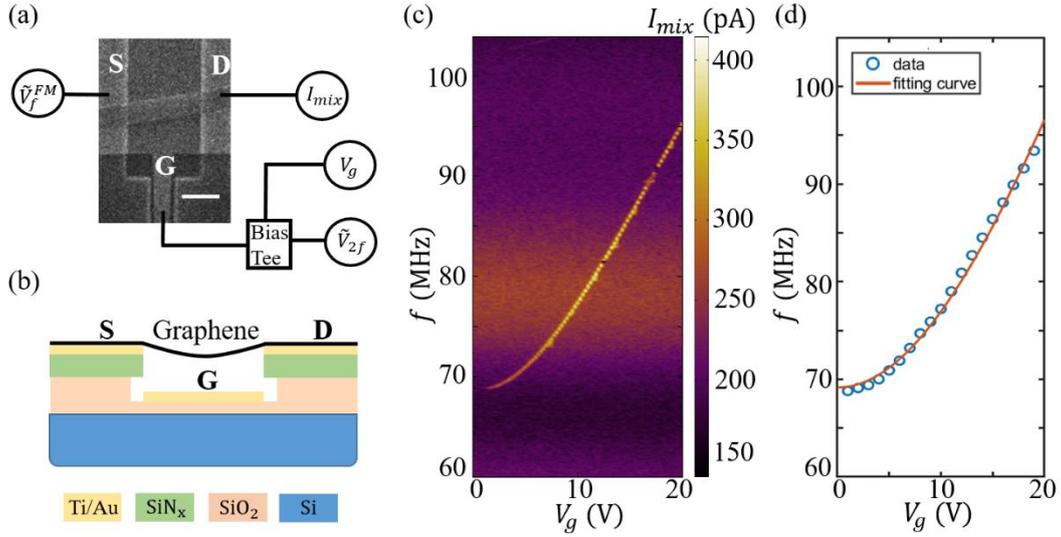

**Fig. 1** (a) Scanning electron microscopy image of the graphene nanomechanical resonator and schematic of the circuit used in the measurement (Scale bar: 1 μm). A frequency-modulated microwave signal $\tilde{V}_f^{FM}$ is applied to the source (S), and the mixing current $I_{mix}$ is detected at the drain (D). The resonator can be electrically tuned by a dc voltage applied to the gate (G). A continuous ac external pumping $\tilde{V}_{2f}$ at a frequency of 2$f$ is applied to the gate through a bias-tee to realize parametric amplification. (b) Schematic of the cross-section of the device. (c) Mixing current spectrum of the nanomechanical resonator. The resonant frequency can be tuned by varying $V_g$. The data is obtained with $P_f^{FM} = -42$ dBm. (d) Fitted curve (red curve) of the resonant frequency as a function of gate voltage, from which the effective mass $m = 5.7 \times 10^{-18}$ kg can be obtained. The data (blue open circles) are extracted from (c).

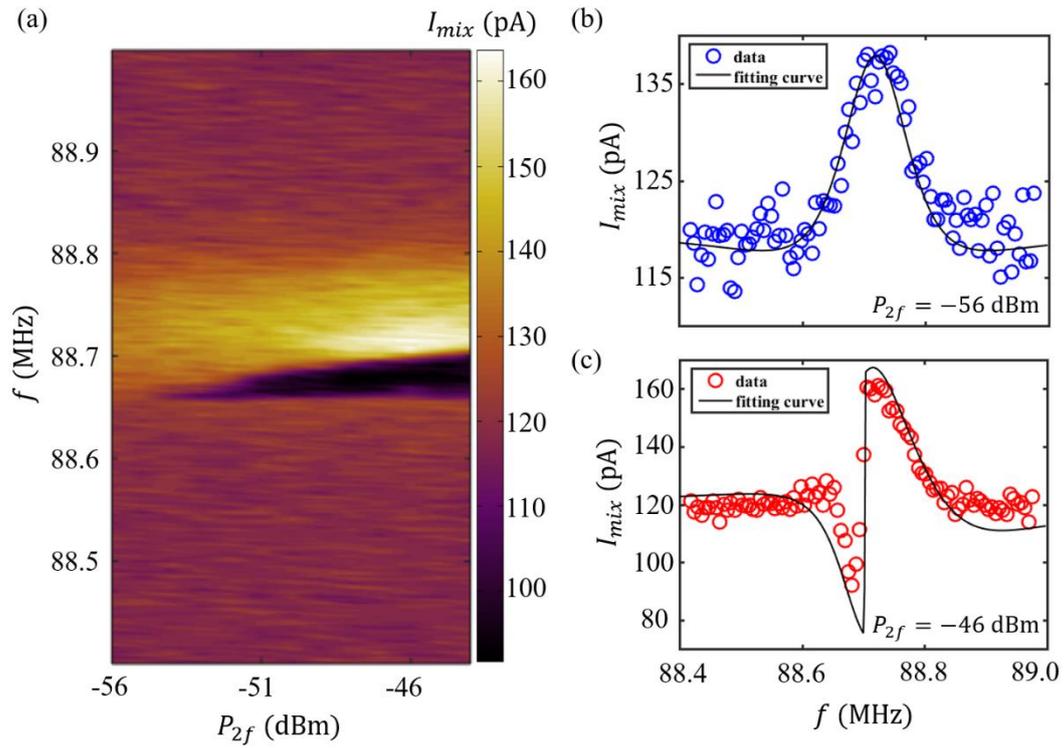

**Fig. 2** (a) Mixing current as a function of driving frequency $f$ and power of external pumping $P_{2f}$. Here, $V_g$=16.5 V. (b)-(c) Mixing current as a function of driving frequency at different pumping powers $P_{2f}$ of −56 dBm and −46 dBm. The open blue (red) circles are experimental data, and the best fitting results are shown by black solid curves.

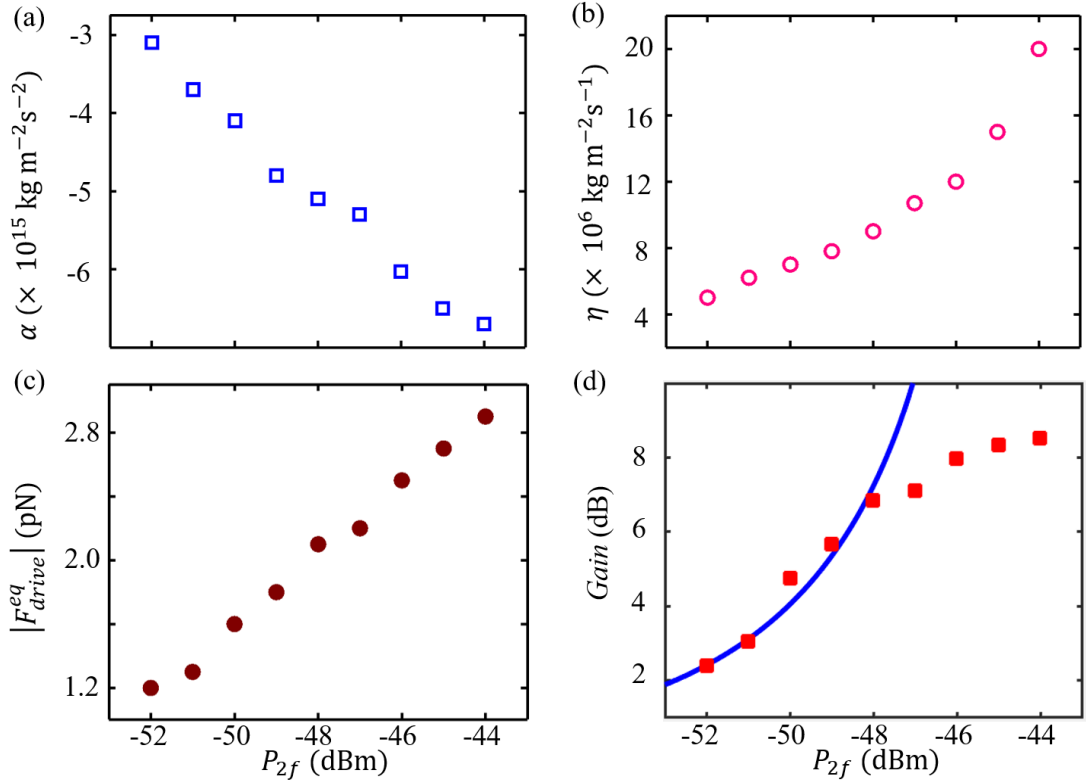

**Fig. 3** (a)-(c) Extracted values of the Duffing force coefficient $\alpha$, nonlinear damping coefficient $\eta$ and equivalent driving force amplitude $|F_{drive}^{eq}|$ as a function of pumping power $P_{2f}$ in the nonlinear regime. (d) Calculated parametric $Gain$ (red solid squares) as a function of $P_{2f}$ at $V_g$=16.5 V. The blue curve is the fitting result using Eq. (7), which fits the data well when the influence of nonlinearity is not significant at lower $P_{2f}$.

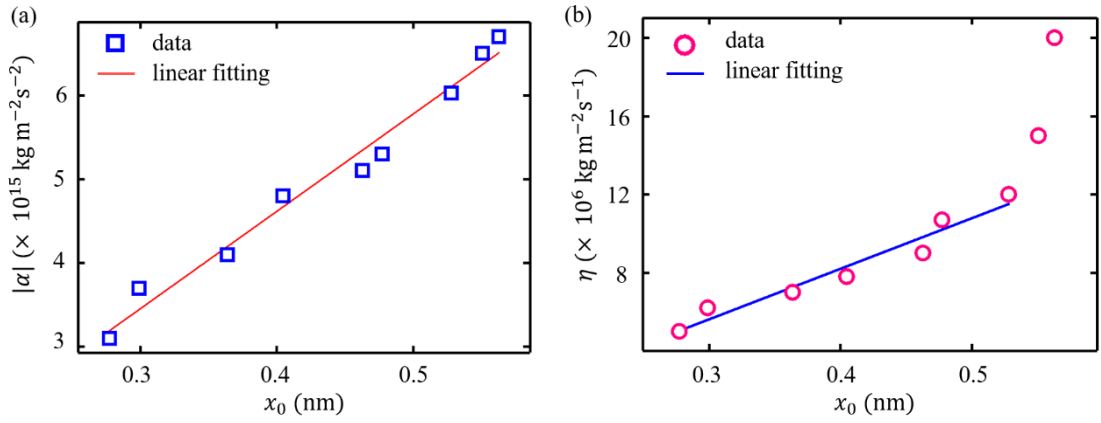

**Fig. 4** (a) The absolute value of the Duffing force coefficient $|\alpha|$ as a function of the vibrational amplitude $x_0$. The red solid line is the linear fitting. (b) The nonlinear damping coefficient $\eta$ as a function of the vibrational amplitude $x_0$, the blue solid line is the linear fitting of the data with $x_0$ smaller than 0.53 nm.

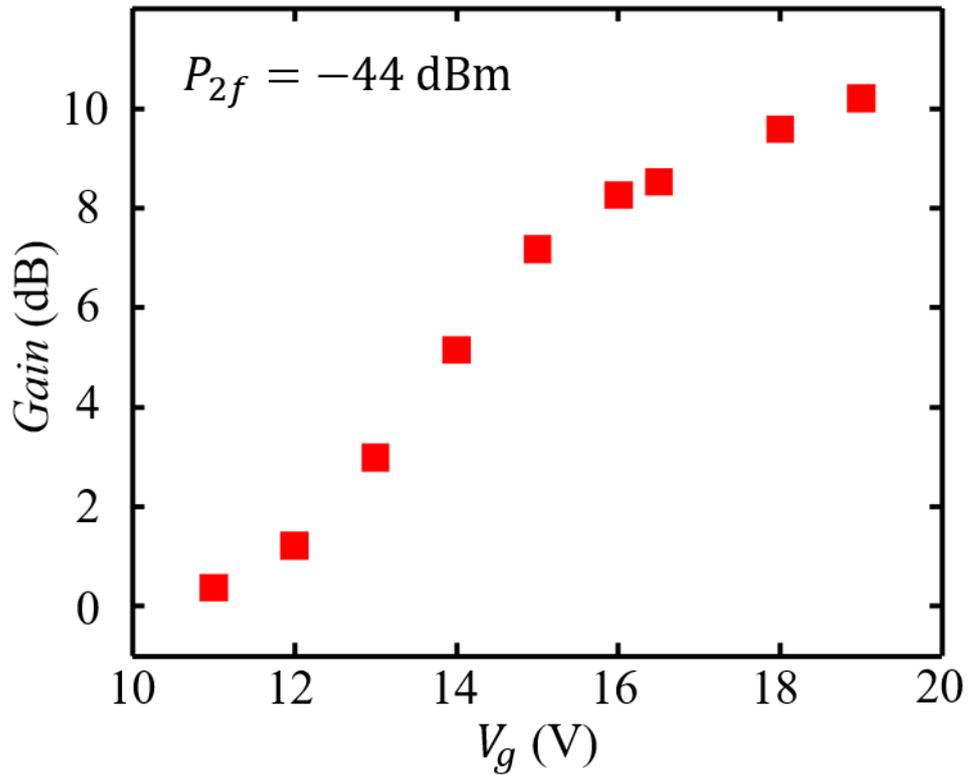

**Fig. 5** Parametric $Gain$ as a function of gate voltage $V_g$ when $P_{2f} = -44$ dBm with $P_f^{FM}$ set to be $-56$ dBm.